\documentclass[showpacs,amsmath,amssymb]{revtex4}

\usepackage{graphics}
\usepackage[dvips]{graphicx}

\def\sqr#1#2{{\vcenter{\hrule height.#2pt\hbox{\vrule width.#2pt height#1pt
\kern#1pt \vrule width.#2pt}\hrule height.#2pt}}}
\def\square{\mathchoice\sqr64\sqr64\sqr{4.2}3\sqr{3.0}3}

\begin{document}

\title{Scalar-tensor cosmological simulations}


\author{M.A. Rodr\'{\i}guez-Meza}
\address{Depto. de F\'{\i}sica, Instituto Nacional de Investigaciones Nucleares,
  Col. Escand\'on, Apdo. Postal 18-1027, 11801 M\'{e}xico D.F. M\'exico \\
  mar@nuclear.inin.mx; http://www.astro.inin.mx/mar
}

\begin{abstract}
We present $N$-body cosmlogical simulations in the framework of the Newtonian
limit of scalar-tensor theories of gravity. The scalar field is described by a modified
Helmholtz equation with a source that is coupled to the standard Poisson equation of Newtonian
gravity. The effective gravitational force is given by two contributions: the standard
Newtonian potential plus a Yukawa potential stemming from massive scalar fields. In particular,
we consider simulations of $\Lambda$CDM models and compute the density and velocity
profiles of the most massive groups found at z=0.
\end{abstract}

\maketitle

\section{Introduction}
In this work we present some preliminary results about the role scalar fields play 
in cosmological simulations, in particular
on the process of large scale structure formation.
Scalar fields have been around for so many years since the
pioneering work of Jordan, Brans, and Dicke\cite{Jordan, BransDicke}.
Nowadays they are considered as a mechanism for inflation\cite{Copeland2004};
the dark matter component of galaxies\cite{Guzman2000};
the quientessence field to explain dark energy in the universe\cite{Axel2004}.
The main goal of this work is to study the large scale structure formation 
where  the usual approach is that the evolution of 
the initial primordial fluctuation energy density fields evolve following 
Newtonian mechanics in an expanding background\cite{Peebles1980}.
The force between particles are the standard Newtonian gravitational force.
Now, we will see that we can introduce the scalar fields by adding a term in this force.
This force will turn out to be of Yukawa type with two parameters 
($\alpha$, $\lambda$)\cite{mar2004}.
For so many years this kind of force, the so called fifth force, 
was thoroughly studied theoretically\cite{Pimentel} 
and many experiments were done to constrain the Yukawa 
parameters\cite{Sudarsky}.
We have also been studying, in the past years, the effects of 
this kind of force on some astrophysical 
phenomena\cite{mar2004,Rodriguez2001,mar2005,jorge2007}.
The Yukawa force comes as a Newtonian limit of a scalar-tensor theory with the scalar 
field non-minimally coupled to gravitation\cite{Helbig} although other alternatives can be 
found\cite{Nusser2005}.
It is our purpose to find the role these scalar fields play on the large scale structure 
formation processes.
We start by discussing the standard LCDM model and the general approach in $N$-body
simulations (See Bertschinger\cite{Bertschinger1998} for details). 
Then, we present the modifications
we need to do to consider the effects of a static scalar field and we show the
results of this theory for the cosmological concordance model of a $\Lambda$CDM 
universe\cite{Heitmann2005}. To perform the simulations we have modified a standard serial 
treecode the author has developed \cite{Gabbasov2006} and the Gadget 1 \cite{Springel2001} 
(see also \url{http://www.astro.inin.mx/mar})
in order to take into account the contribution of the Yukawa potential.

\section{Evolution equations for a $\Lambda$CMD universe}
\subsection{Newtonian approximation}
The study of large-scale formation in the universe is greatly simplified by the fact that a
limiting approximation of general relativity, Newtonian mechanics, applies in a region
small compared to the Hubble length $cH^{-1}$ ($cH_0^{-1}\approx 3000 h^{-1}$ Mpc, where 
$c$ is the speed of light, $H_0=100 h$ km/s/Mpc, 
is Hubble's constant and $h\approx (0.5-1)$), and large compared to the Schwarzschild
radii of any collapsed objects. The rest of the universe affect the region only through a tidal field.
The length scale $cH_0^{-1}$ is of the order of the largest scales currently accessible in 
cosmological observations and $H_0^{-1} \approx 10^{10}h^{-1}$ yr 
characterizes the evolutionary time scale of the universe.

The Newtonian approximatiion can fail at much smaller $R$ if the region includes a 
compact object like a neutron star or black hole, but one can deal with this by noting that at
distances large compared to the Schwarzschild radius the object acts like an ordinary 
Newtonian point mass. It is speculated that in nuclei of galaxies there might be black holes
as massive as $10^9$ M$_\odot$, Schwarzschild radius $\sim 10^{14}$ cm. If this is an upper
limit, Newtonian mechanics is a good approximation over a substancial range of scales, 
$10^{14}$ cm $\ll r \ll 10^{28}$ cm.

\subsection{General Scalar-tensor theory}
Let us consider a typical scalar--tensor theory given by the following
Lagrangian
\begin{equation}\label{EqSTTLagrangian}
{\cal L} = \frac{\sqrt{-g}}{16\pi} \left[ -\phi R + \frac{\omega(\phi)}{\phi}
	(\partial \phi)^2 - V(\phi) \right] + {\cal L}_M(g_{\mu\nu}) \; ,
\end{equation}
Here $g_{\mu\nu}$ is the metric,
${\cal L}_M(g_{\mu\nu})$ is the matter Lagrangian and $\omega(\phi)$ and
$V(\phi)$ are arbitrary functions of the scalar field.
From Lagrangian (\ref{EqSTTLagrangian}) we get the gravitational equations,
\begin{eqnarray}
R_{\mu\nu} - \frac{1}{2} g_{\mu\nu} R &=& \frac{1}{\phi}
\left[ 8 \pi T_{\mu\nu} + \frac{1}{2} V g_{\mu\nu}
+ \frac{\omega}{\phi} \partial_\mu \phi \partial_\nu \phi
\right. \nonumber \\
&& \left. -\frac{1}{2} \frac{\omega}{\phi}(\partial \phi)^2 g_{\mu\nu}
+ \phi_{;\mu\nu} - g_{\mu\nu} \, \square \phi \frac{\mbox{}}{\mbox{}}
\right] \; , 
\end{eqnarray}
and the scalar field equation
\begin{equation}
\square \phi + \frac{\phi V' - 2V}{3+2\omega} = \frac{1}{3+2\omega} \left[
	8\pi T -\omega' (\partial \phi)^2 \right] \, ,
\end{equation}
where $()' \equiv \frac{\partial }{\partial \phi}$. 
The gravitational constant is now contained in $V(\phi)$ and the scalar field get a mass $m_{SF}$.

\subsection{Newtonian approximation of STT}
In the present study, however, we want to consider the influence
of scalar fields in the limit of a static STT, and therefore we need to describe
the theory in its Newtonian approximation, that is, where gravity 
and the scalar fields are weak (and time independent) and velocities of stars are
non-relativistic.  We expect to have small deviations of
the scalar field around the background field, defined here as
$\langle \phi \rangle$ and can be understood as the scalar field beyond all matter.
If one defines the perturbations $\phi = \langle \phi \rangle + \bar{\phi}$ and
$g_{\mu\nu} = \eta_{\mu\nu} + h_{\mu\nu}$,
where $\eta_{\mu\nu}$ is the Minkowski metric, the Newtonian approximation
gives \cite{Helbig}
\begin{eqnarray}
R_{00} = \frac{1}{2} \nabla^2 h_{00} &=& \frac{G_N}{1+\alpha} 4\pi \rho
- \frac{1}{2} \nabla^2 \bar{\phi}  \; ,
\label{pares_eq_h00}\\
  \nabla^2 \bar{\phi} - m_{SF}^2 \bar{\phi} &=& - 8\pi \alpha\rho \; ,
\label{pares_eq_phibar}
\end{eqnarray}
we have set $\langle\phi\rangle=(1+\alpha)/G_N$ 
and $\alpha \equiv 1 / (3 + 2\omega)$.  In the above expansion we have set
the cosmological constant term equal to zero, since on galactic
scales its influence should be negligible.  We only consider the
influence of dark matter due to the boson field of mass $m_{SF}$ governed by
Eq.\ (\ref{pares_eq_phibar}), that is the modified Helmholtz equation.
Equations (\ref{pares_eq_h00}) and (\ref{pares_eq_phibar}) represent
the Newtonian limit of STT with arbitrary potential $V(\phi)$ and function
$\omega(\phi)$ that where Taylor expanded around $\langle\phi\rangle$.
The resulting equations are then distinguished by the constants
$G_N$, $\alpha$, and $\lambda=h_P/m_{SF}c$. Here $h_P$ is Planck's constant.

Note that Eq. (\ref{pares_eq_h00}) can be cast as a Poisson equation for
$\psi \equiv (1/2) ( h_{00} + \bar{\phi}/ \langle \phi\rangle)$, 
\begin{equation}
\nabla^2 \psi = 4\pi \frac{G_N}{1+\alpha} \rho \; . \label{pares_eq_psi}
\end{equation}

The next step is to find solutions for this new Newtonian potential given 
a density profile, that is, to find the so--called potential--density pairs. 
General solutions to Eqs. (\ref{pares_eq_phibar}) and (\ref{pares_eq_psi})
can be found in terms of the corresponding Green functions,
and the new Newtonian potential is
\begin{eqnarray}
\Phi_N  \equiv \frac{1}{2} h_{00}
&=& - \frac{G_N}{1+\alpha} \int d{\bf r}_s
\frac{\rho({\bf r}_s)}{|{\bf r}-{\bf r}_s|} \nonumber \\
&& -\alpha \frac{G_N}{1+\alpha} \int d{\bf r}_s \frac{\rho({\bf r}_s)
{\rm e}^{- |{\bf r}-{\bf r}_s|/\lambda}}
{| {\bf r}-{\bf r}_s|} + \mbox{B.C.} \label{pares_eq_gralPsiN}
\end{eqnarray}
The first term of Eq. (\ref{pares_eq_gralPsiN}), given by $\psi$, is the
contribution of the usual Newtonian gravitation (without scalar
fields), while information about the scalar field is contained in the
second term, that is, arising from the influence function determined by the
modified Helmholtz Green function, where the coupling $\omega$ ($\alpha$) enters
as part of a source factor.

\subsection{Cosmological evolution equations using a static STT}
To simulate cosmological systems,  the expansion of the universe has to be
taken into account.
Also, to determine the nature of the cosmological model we need to determine
 the composition of the
universe, i. e., we need to give the values of $\Omega_i$ for each component $i$, 
taking into
in this way all forms of energy densities that exist at present.

If a particular kind of energy density is described by an equation of state of the form
$p=w \rho$, where $p$ is the pressure and $w$ is a constant, then the equation for energy
conservation in an expanding background, $d(\rho a^3)=-pd(a^3)$, can be integrated to
give $\rho \propto a^{-3(1+w)}$. Then, the Friedmann equation for the expansion factor $a(t)$
is written as
\begin{equation}
\frac{\dot{a}^2}{a^2} = H_0^2 \sum_i \Omega_i \left(\frac{a_0}{a}\right)^{3(1+w_i)} - \frac{k}{a^2}
\end{equation}
where $w_i$ characterizes equation of state of specie $i$. The most familiar forms of energy
densities are those due to pressureless matter with $w_i=0$ (that is, nonrelativistic matter
with rest-mass-energy density $\rho c^2$ dominating over the kinetic-energy density
$\rho v^2/2$) and radiation with $w_i=1/3$.  The density parameter contributed today
by visible, nonrelativistic, baryonic matter in the universe is $\Omega_B \approx (0.01-0.2)$
and the density parameter that is due to radiation is $\Omega_R \approx 2\times 10^{-5}$.
In this work we will consider a model with only two energy density contribution.
One which is a pressureless and
nonbaryonic dark matter  with $\Omega_{DM} \approx 0.3$ that does not couple with radiation.
Other, that will be a cosmological constant contribution $\Omega_\Lambda \approx 0.7$
with and equation of state $p =-\rho$. The above equation for $a(t)$ becomes
\begin{equation}
\frac{\dot{a}^2}{a^2} = H_0^2 
\left[ 
\Omega_{DM} \left(\frac{a_0}{a}\right)^{3} +  \Omega_\Lambda
\right]
- \frac{k}{a^2}
\end{equation}

Here, we employ a cosmological model with a static scalar field which is consistent with the 
Newtonian limit given by Eq. (\ref{pares_eq_gralPsiN}).
Thus, the scale factor, $a(t)$,  is given by the following Friedman model,
\begin{equation} \label{new_friedman}
a^3 H^2= H_{0}^{2} \left[\frac{\Omega_{m0} +  \Omega_{\Lambda 0} \, a^3}{1+\alpha} 
+  \left(1-\frac{\Omega_{m 0}+\Omega_{\Lambda 0}}{1+\alpha} \right) \, a  \right]
\end{equation}
where $H=\dot{a}/a$,  $\Omega_{m 0}$ and $\Omega_{\Lambda 0}$ 
are the matter and energy density evaluated at present, respectively.   
We notice that the source of the cosmic evolution is deviated by the term 
$1+\alpha$ when compared to the standard Friedman-Lemaitre 
model. Therefore, it is convenient to define a new density parameter by 
$\Omega_i^{\alpha} \equiv \Omega_i/(1+\alpha)$. This new density parameter is such that 
$\Omega_m^{\alpha} + \Omega_\Lambda^{\alpha} =1$, 
which implies a flat universe, and this shall be assumed 
in our following computations, where we consider 
$(\Omega_m^{\alpha}, \Omega_\Lambda^{\alpha}) = (0.3, 0.7) $.  For positive values 
of $\alpha$, a flat cosmological model demands to have a factor $(1+\alpha)$ more energetic 
content ($\Omega_m$ and $ \Omega_\Lambda$) than in standard cosmology. 
On the other hand, for negative values of  
$\alpha$ one needs a factor $(1+\alpha)$  less $\Omega_m$ 
and $ \Omega_\Lambda$ to have a flat universe.  To be consistent 
with the CMB  spectrum and structure formation numerical 
experiments, cosmological constraints must be applied on $\alpha$ in order for it to 
be within the range $(-1,1)$ \cite{Nagata2002,Nagata2003,Shirata2005,Umezu2005}.  

In the Newtonian limit of STT of gravity, 
the Newtonian motion equation  for a particle $i$ is written as
\begin{equation} \label{eq_motion}
\ddot{\mathbf{x}}_i + 2\, H \, \mathbf{x}_i = 
-\frac{1}{a^3} \frac{G_N}{1+\alpha} \sum_{j\ne i} \frac{m_j (\mathbf{x}_i-\mathbf{x}_j)}
{|\mathbf{x}_i-\mathbf{x}_j|^3} \; F_{SF}(|\mathbf{x}_i-\mathbf{x}_j|,\alpha,\lambda)
\end{equation}
where $\mathbf{x}$ is the comovil coordinate, and  the sum includes all  
periodic images of particle $j$,  and $F_{SF}(r,\alpha,\lambda)$ is
\begin{equation}
F_{SF}(r,\alpha,\lambda) = 1+\alpha \, \left( 1+\frac{r}{\lambda} \right)\, e^{-r/\lambda}
\end{equation}
which,  for small distances compared to $\lambda$,  is 
$F_{SF}(r<\lambda,\alpha,\lambda) \approx 1+\alpha \, \left( 1+\frac{r}{\lambda} \right)$ and, 
for long 
distances, is  $F_{SF}(r>\lambda,\alpha,\lambda) \approx 1$, as in Newtonian physics. 

We now analyze the general effect that the constant $\alpha$ has on the dynamics.  
The role of $\alpha$ in our approach is
as follows.   On one hand, to construct a flat model  
we have set the condition $\Omega_m^{\alpha} + \Omega_\Lambda^{\alpha} =1$, which 
implies  having $(1+\alpha)$ times the energetic content of the standard $\Lambda$CDM model.
 This essentially means that
we have an increment by a factor of  $(1+\alpha)$ times the amount of matter, 
for positive values of  $\alpha$, or a 
reduction of the same factor for negative values of  $\alpha$. Increasing or reducing 
this amount of matter affects 
the matter term on the  r.h.s. of the equation of 
motion (\ref{eq_motion}), but the amount affected cancels out with the term $(1+\alpha)$ in the 
denominator of  (\ref{eq_motion}) stemming from the new Newtonian potential. 
On the other hand, the factor $F_{SF}$ augments (diminishes) for positive (negative)  
values of $\alpha$ for small distances compared to  $\lambda$, resulting in more (less) structure formation for positive (negative) values of $\alpha$ compared to the $\Lambda$CDM model.  
For $r\gg \lambda$ the dynamics is essentially Newtonian.

\section{Results}
In this section, we present results of the cosmological simulations of a $\Lambda$CDM universe
with and without SF contribution. We consider the {\em smal box}
initial condition in the Cosmic Data Bank web page (\url{http://t8web.lanl.gov/people/heitmann/test3.html}). The initial condition use a box 90 Mpc  size and $256^3$ particles. It
is somewhat small as a representative simulation due to lack of large-scale power but straddle
a representative range of force and mass resolutions for state-of-the-art large scale structure
simulations designed to study power spectra, halo mass functions, weak lensing, and so on.

The initial linear power spectrum was generated using the fitting formula by Klypin \& 
Holtzman\cite{KlypinHoltzman1997} for the transfer function. This formula is a slight variation
of the common BBKS fit\cite{Bardeen1986}. It includes effects from baryon suppression but no 
baryonic oscillations. We use the standard Zel'dovich approximation\cite{Zeldovich1970} 
to provide the initial particle displacement off a uniform grid and to assign initial particle
velocities. The starting redshift is $z_{in}=50$ and we choose the following cosmology:
$\Omega_{m}=0.314$ (where $\Omega_{DM}$ includes cold dark matter and baryons),
$\Omega_{b}=0.044$, $\Omega_{\Lambda}=0.686$, $H_0=71$ km/s/Mpc, $\sigma_8=0.84$,
and $n=0.99$. These values are in concordance with measurements of cosmological
parameters by WMAP\cite{Spergel2003}. The simulations we present here use an initial condition
with only 
$723,925$ particles,
obtained from the original initial condition by a reduction procedure based on a tree scheme.
This implies that 
particle masses are in the order of $1.0\times 10^{10}$ M$_\odot$. 
The individual softening length
was 20 kpc$/h$. These choices of softening length are consistent
with the mass resolution set by the number of particles.

We now present results for the $\Lambda$CDM model previously described. 
Because the visible component is the smaller one and given our interest to
test the consequences of including a SF contribution to the evolution equations,
our model excludes gas particles, but all its mass has been added to the dark matter. 
We restrict the values of $\alpha$ to the interval $(-1,1)$ 
  \cite{Nagata2002,Nagata2003,Shirata2005,Umezu2005}  and  use $\lambda=5$ Mpc$/h$, since 
this scale turns out to be an intermediate scale between the size of the clump groups and 
the separation of  the formed groups.

\begin{figure}
\begin{minipage}{6in}
\begin{center}
\includegraphics[width=2.75in]{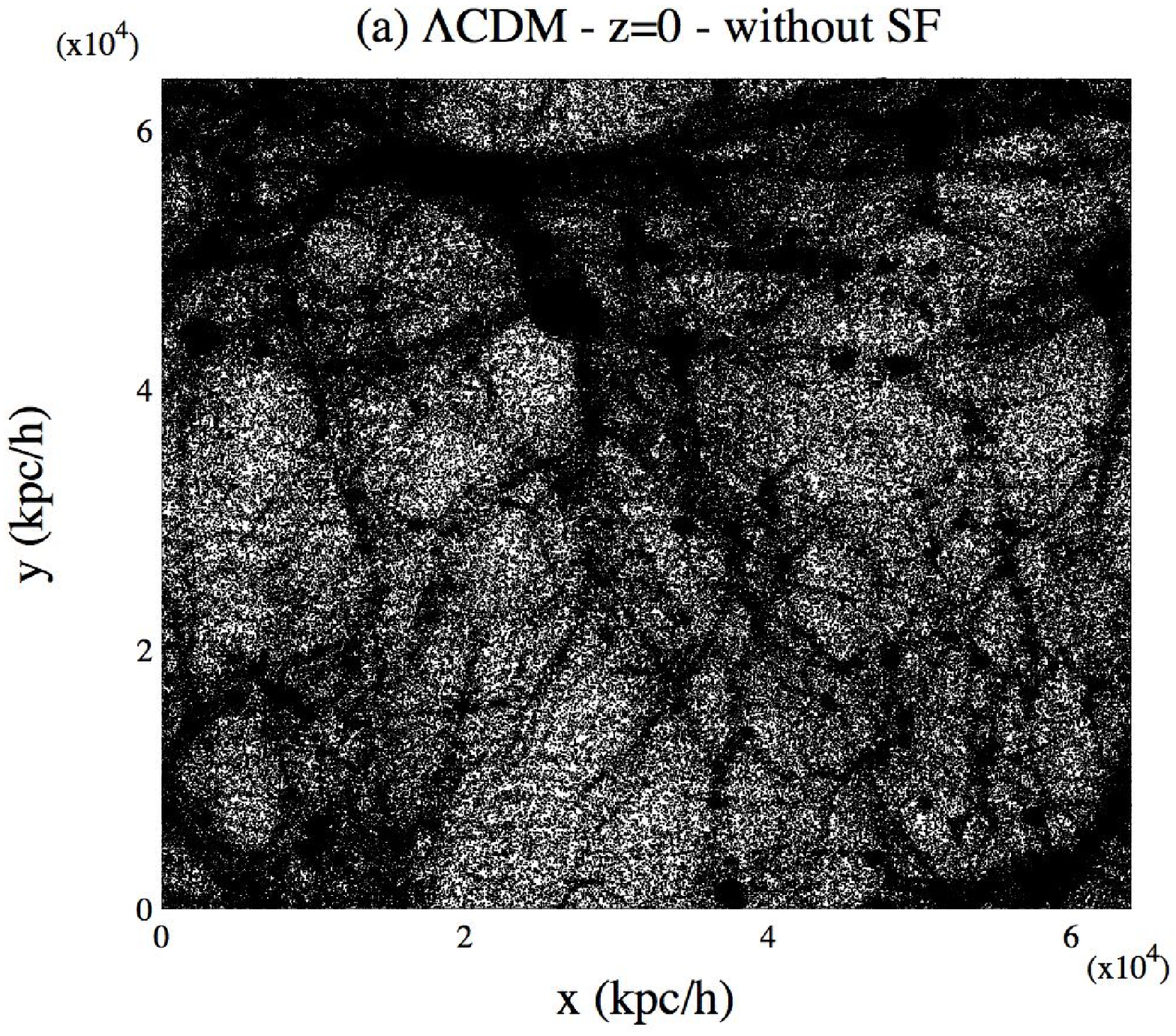}
\includegraphics[width=2.75in]{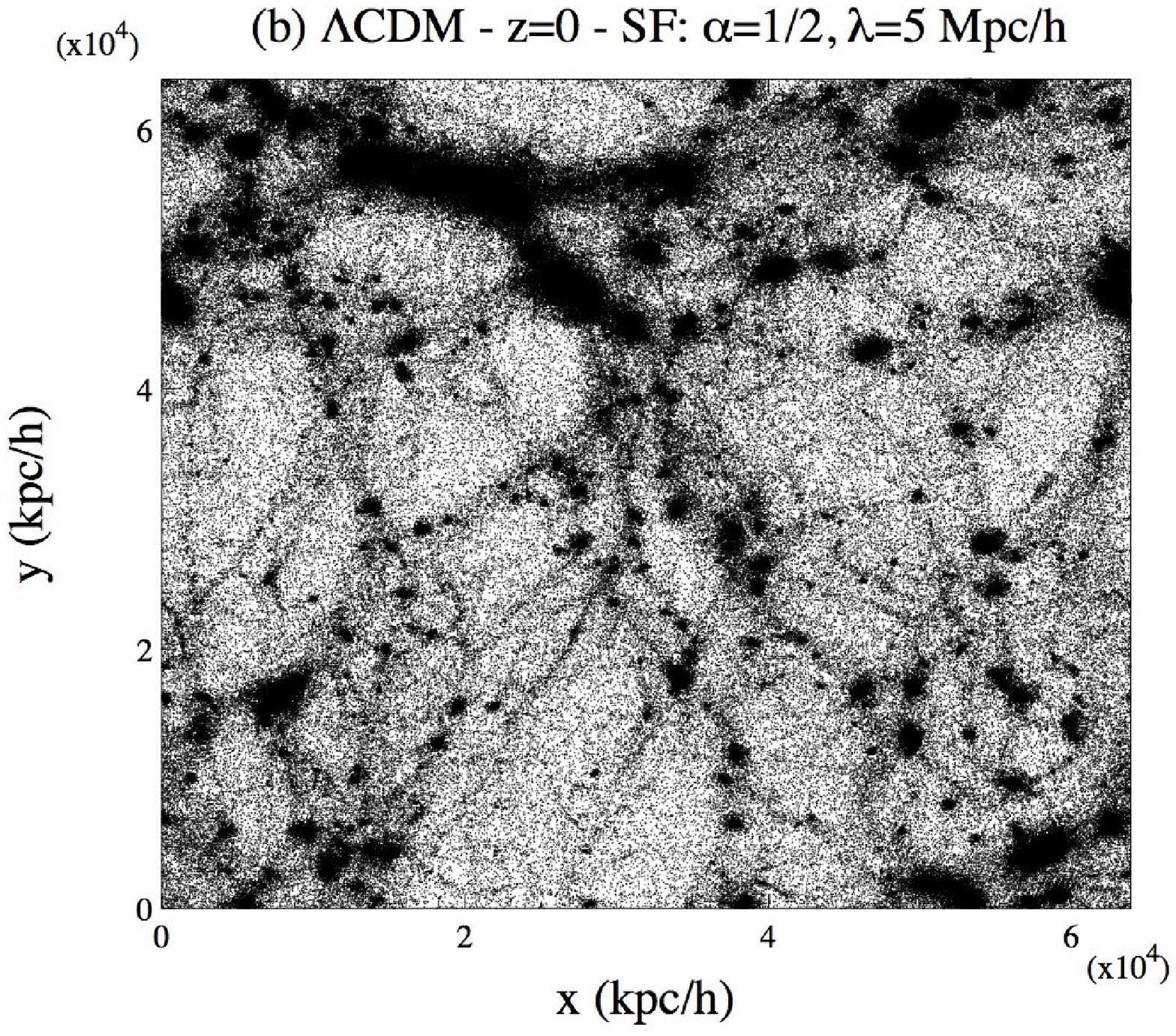}
\includegraphics[width=2.75in]{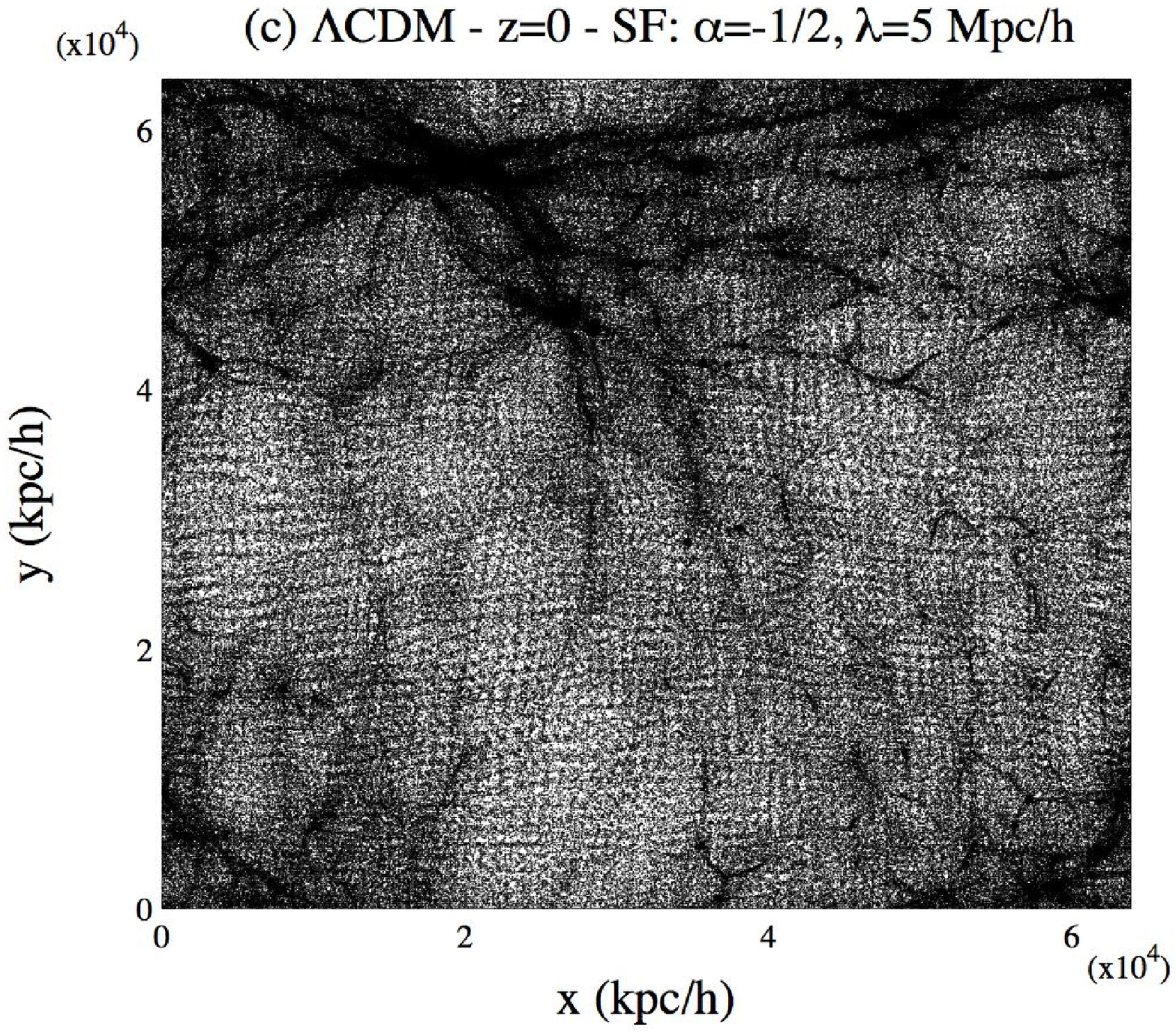}
\includegraphics[width=2.75in]{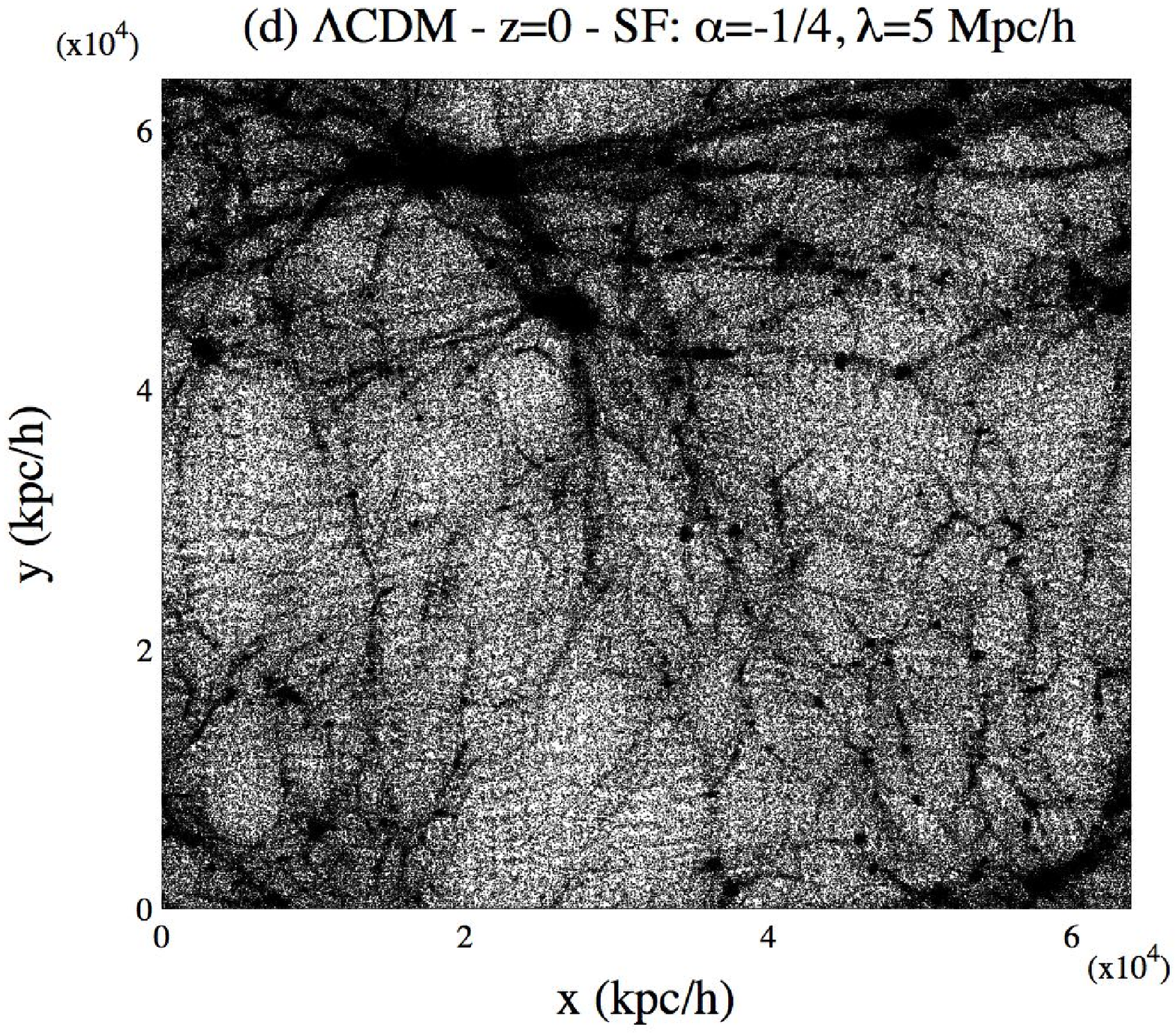}
\end{center}
\end{minipage}
\caption{$x$--$y$ snapshots at $z=0$ of a $\Lambda$CDM universe.  See text for details. }
\end{figure}
In Fig. 1 we show $x$--$y$ snapshots at redshift $z=0$ of our  $\Lambda$CDM model. 
Fig. 1 (a) presents the standard case without SF, i.e., the interaction between bodies  is through
the standard Newtonian potential.
In (b) we show the case with $\alpha=1/2$, $\lambda=5$ Mpc$/h$.
In (c)  $\alpha=-1/2$, $\lambda=5$ Mpc$/h$.
In (d) $\alpha=-1/4$, $\lambda=5$ Mpc$/h$.  
One notes clearly how the SF modifies the matter  structure of the system. The most
dramatic cases are (b) and (c) where we have used 
$\alpha=1/2$ and $\alpha=-1/2$, respectively. 
Given the argument  at the end of last section, in the case of (b), for $r \ll \lambda$,
the effective gravitational pull has been  augmented by a factor of $3/2$, 
in contrast to case (c) where it has diminished  by a factor of 1/2; in model (d) the pull 
diminishes only by a factor of 3/4. That is why one observes for $r < \lambda$ more structure 
formation in (b), less in (d), and lesser in  model (c).  
The effect is  then, for a growing positive $\alpha$, 
to speed up  the growth of perturbations, then of halos and then of clusters, whereas negative 
$\alpha$ values ($\alpha \rightarrow -1$) tend to slow down the growth. 

\begin{figure}
\includegraphics[width=3.1in]{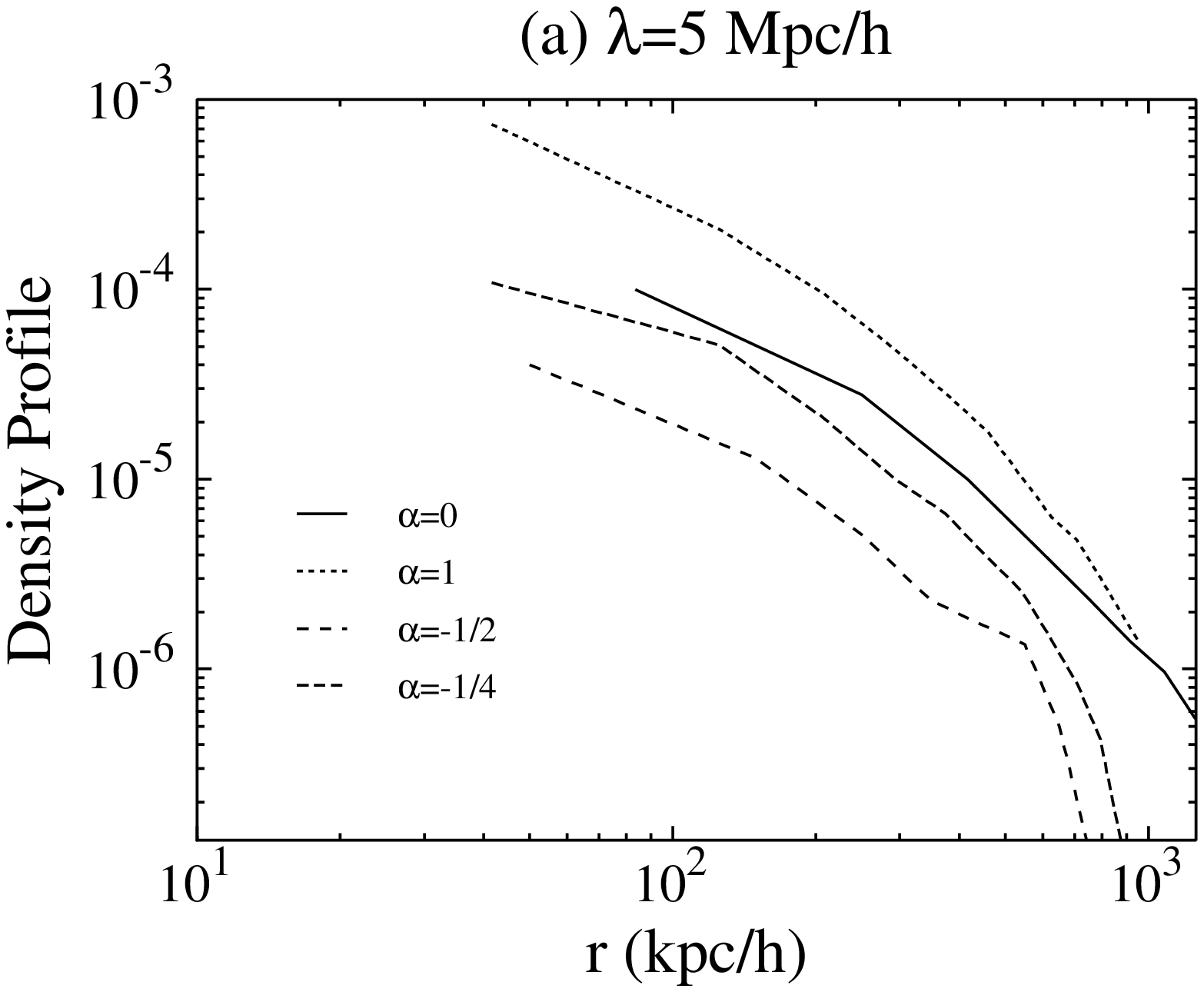}
\includegraphics[width=3.1in]{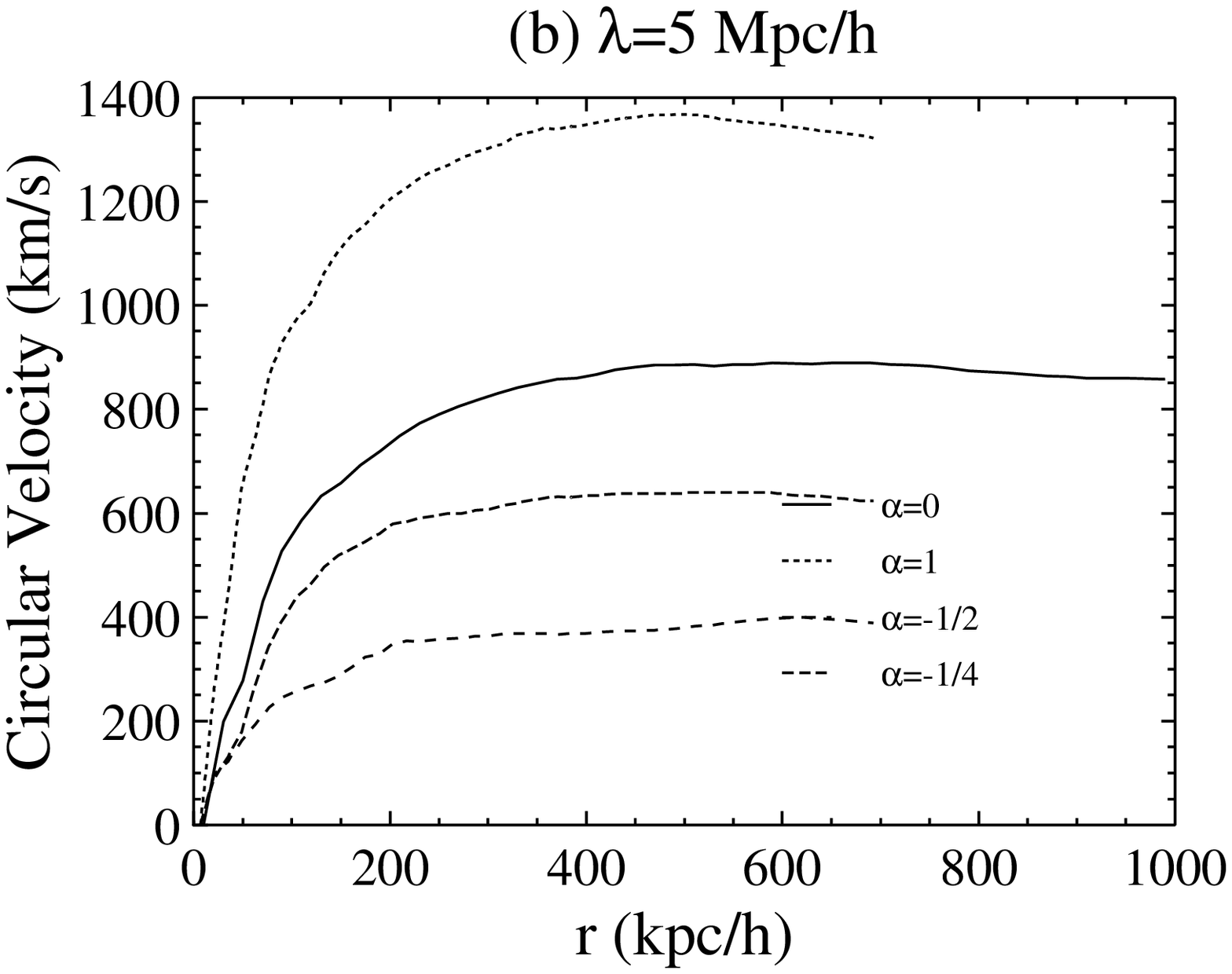}
\caption{(a) Density profiles for one of the most massive groups 
at $z=0$ of a $\Lambda$CDM universe. 
 The group is located approximately at $x=14$ Mpc$/h$, $y=57.5$ Mpc$/h$ in Fig 1(a). 
Vertical scale is in units of $\rho_0=10^{10} \mbox{M}_\odot h^{-1} / (h^{-1}\mbox{kpc})^3$.
(b) The corresponding circular velocity.}
\end{figure}
Next, we found the groups in the system using a friend-of-friend algorithm
and select one of the most massive ones. The chosen group is located
approximately at $x=14$ Mpc$/h$, $y=57.5$ Mpc$/h$.
The group was analyzed by obtaining their density profiles (Fig. 2(a))
and circular velocities (Fig. 2(b)).
The  more cuspy case is for $\alpha = 1/2$ and
the less cuspy is for $\alpha=-1/2$. 
The circular velocity curves where computed using 
$v_c^2=G_N M(r)/r$. The case with $\alpha=1/2$ corresponds to higher values of $v_c$, 
since this depends on how much accumulated mass there is at a distance $r$ and this
is enhanced by the factor 
$F_{SF}$ for positive values of $\alpha$.

\section{Conclusions}

In general, we can say that even though we have done first numerical
simulations using non-minimally coupled SF, the analysis we have done is insufficient to give us a clear
conclusions on the role played by SF in the large-scale structure formation process.
We will need to do a systematic study of the evolution of the two-point correlation function which
is a mesure of galaxy clustering.
We also will need to compute the mass power
spectrum and velocity dispersions of the halos. Therefore, we will be able make 
sistematic comparisons with observations. However, and in favor of the model, the theoretical
scheme we have used 
is compatible with local observations because we have defined
the background field constant 
$<\phi>  =  G_{N}^{-1} (1+\alpha)$.  A direct consequence of  
the approach is that  the amount of matter (energy) has to be increased 
for positive values of $\alpha$ and diminished  for negative values of $\alpha$ 
with respect to the standard $\Lambda$CDM model 
in order to have a flat cosmological model. Quantitatively, our model demands to 
have $\Omega/ (1+\alpha) =1$ and this changes the amount of dark matter and 
energy of the model for a flat cosmological model, as assumed.   
The general gravitational effect is that  the interaction including  the SF changes by a factor 
$F_{SF}(r,\alpha,\lambda) \approx 1+\alpha \, \left( 1+\frac{r}{\lambda} \right)$ for $r<\lambda$ in 
comparison with the Newtonian case. Thus, for $\alpha >0$ the growth of structures speeds up  
in comparison with the Newtonian case.  For the   $\alpha <0$ case the effect is to diminish 
the formation of structures.  For $r> \lambda$ the dynamics is essentially Newtonian.

\bigskip
{\it Acknowledgments}
This work was supported by CONACYT, grant number I0101/131/07 C-234/07, IAC collaboration. 
The simulations were performed in the UNAM HP cluster {\it Kan-Balam}.

\end{document}